\documentclass[twocolumn]{revtex4-1}
\usepackage{helvet}

\usepackage[latin9]{inputenc}
\usepackage{geometry}
\usepackage{xcolor}
\usepackage{pdfcolmk}
\usepackage{amsmath}
\usepackage{graphicx}
\PassOptionsToPackage{normalem}{ulem}
\usepackage{ulem}

\makeatletter

\providecolor{lyxadded}{rgb}{0,0,1}
\providecolor{lyxdeleted}{rgb}{1,0,0}

\makeatother

\begin{document}

\title{Independent Noise Synchronizing Networks of Oscillator Networks}

\author{John Hongyu Meng$^{1}$ and Hermann Riecke$^{1,2,+}$ }

\address{$^{1}$Engineering Sciences and Applied Mathematics, Northwestern
University, Evanston, IL 60208, USA}

\address{$^{2}$Northwestern Institute on Complex Systems, Northwestern University,
Evanston, IL 60208, USA }

\address{$^{+}$corresponding author}
\begin{abstract}
Oscillators coupled in a network can synchronize with each other to
yield a coherent population rhythm. If multiple such networks are
coupled together, the question arises whether these rhythms will synchronize.
We investigate the impact of noise on this synchronization for strong
inhibitory pulse-coupling and find that increasing the noise can synchronize
the population rhythms, even if the noisy inputs to different oscillators
are completely uncorrelated. Reducing the system to a phenomenological
iterated map we show that this synchronization of the rhythms arises
from the noise-induced phase heterogeneity of the oscillators. The
synchronization of population rhythms is expected to be particularly
relevant for brain rhythms. 
\end{abstract}
\maketitle
The synchronization of coupled oscillators has been studied extensively
\cite{DoeBul14,RodKur16}. Classical examples of technologically important
synchronization include arrays of microwave oscillators \cite{YoCo91}
or lasers \cite{BrJo05}, networks of Josephson junctions \cite{WiCo96}
and optomechanical oscillators \cite{ZhaLip12}. Biological systems
in which synchronization plays a central functional role include pacemaker
cells in the heart \cite{MiMa87} and in the suprachiasmatic nucleus
of the brain, which controls the circadian rhythm \cite{LiWe97}.
Various types of rhythmic, coherent activity of large ensembles of
neurons have been observed in many brain regions \cite{Wa10}. What
functional role they may play in the information processing performed
by the brain is still under debate \cite{BuzSch15,Fri15}.

Noise typically counteracts the synchronization of oscillators. Only
if the noise driving different oscillators is sufficiently correlated
does an increase in the noise level lead to synchronization \cite{Pi84,ZhKu02,TeTa04}.
This noise-induced synchronization does not require any coupling between
the oscillators; it essentially reflects the transfer of noise correlations
from the input to the output \cite{ShJo08,AboErm11}.

If the oscillators in a network are sufficiently strongly synchronized
due to their coupling, the whole network of oscillators can be thought
of as a single composite network oscillator undergoing rhythmic population
activity. If multiple such networks, each supporting its own rhythm,
are coupled together, the question arises under what conditions such
population rhythms phase-lock or synchronize and how that synchronization
is affected by noise. Does the composite nature of the network oscillators
introduce aspects that are not found in coupled individual oscillators?

Brain rhythms constitute an important class of population dynamics
of oscillator networks. Among these the widely observed $\gamma$-rhythm
(30-100 Hz) has been studied particularly extensively \cite{BuzWan12}.
It typically arises either from the inhibitory interaction among interneurons
(`ING-rhythm') or the reciprocal interaction between excitatory and
inhibitory neurons (`PING-rhythm') \cite{WhTr00,BrWa03,BoKo03,TiSe09}.
Importantly, $\gamma$-rhythms can arise simultaneously in different
brain areas and rhythms in different areas may or may not be coherent
with each other \cite{BosFri12}. Even within a single brain area
they can differ in frequency and phase \cite{NeHa03,KaLa10}. Experiments
show that the degree of coherence of $\gamma$-rhythms in different
brain areas can depend on the attentional state of the animal \cite{GrGo09,BosFri12,RobDeW13}
and it has been suggested that this coherence may play an important
role in the communication between these areas \cite{Fri15}.

\begin{figure}
\centering{}\includegraphics[width=0.95\linewidth]{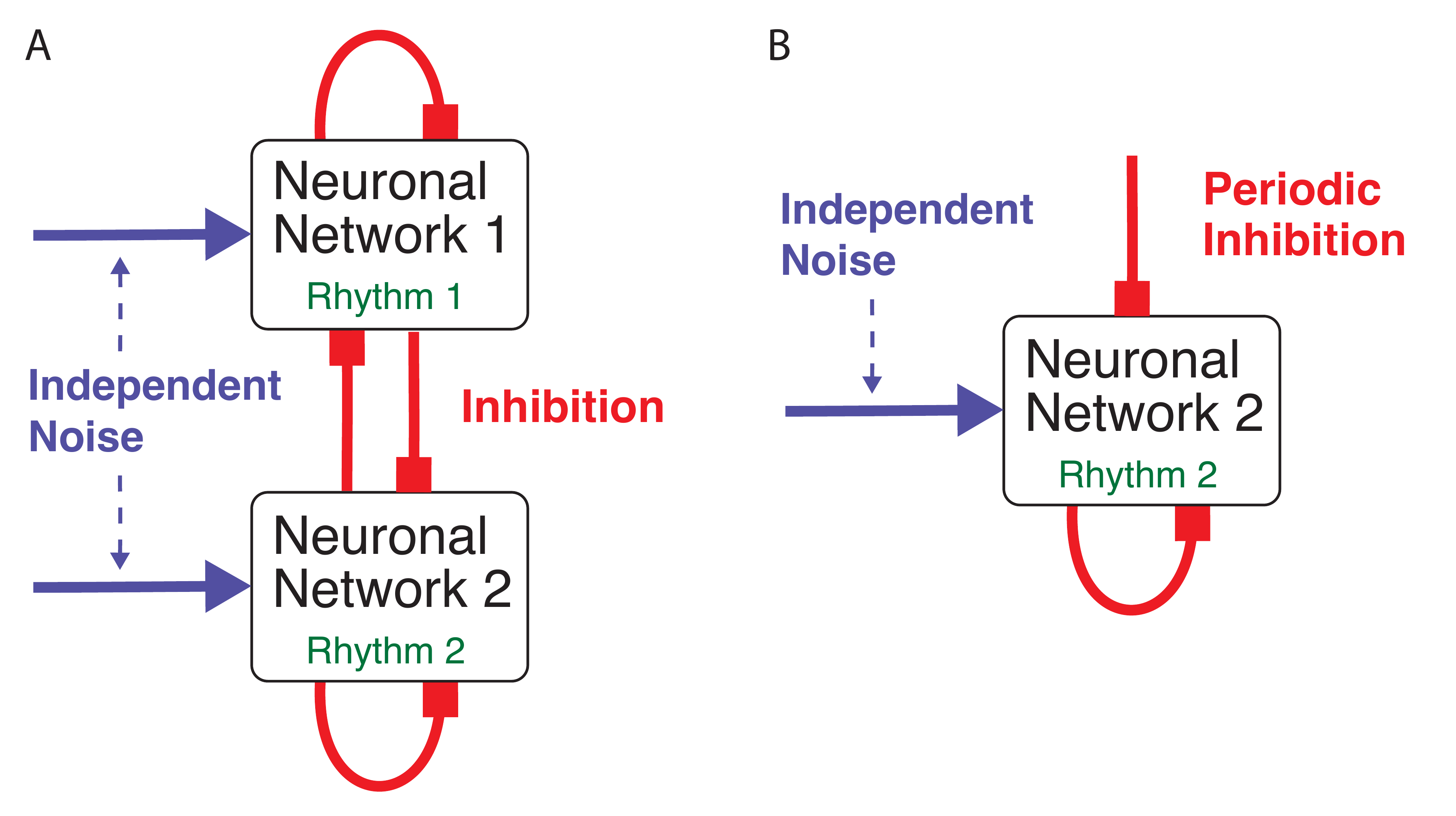}\caption{A) Two coupled inhibitory networks exhibiting separate $\gamma$-rhythms
driven by independent noise. B) Simplified system with network 1 replaced
by time-periodic inhibition.}
\label{fig:figure1} 
\end{figure}

This paper is motivated by i) the wide-spread occurence of modular
networks \cite{RaSo02,BuSp09}, which can be considered as coupled
networks, ii) the observation of coherence of $\gamma$-rhythms across
different brain areas \cite{BosFri12}, and iii) the appearance of
multiple, different $\gamma$-rhythms in a single brain area \cite{NeHa03,KaLa10}
that is likely to exhibit modular structure as a result of learning
via structural plasticity \cite{ChoRie12,SaiLle16}. To investigate
the role noise can play in the interaction of such rhythms we consider
coupled neuronal networks that each support their own $\gamma$-rhythm
and that interact via inhibitory pulses (Fig.\ref{fig:figure1}A).
The synchronization of modular networks has been studied using the
master stability function approach \cite{PaLa06,SoOt07} and for diffusively
coupled phase oscillators within the framework of the Kuramoto model
\cite{OhRh05,ArDi08,GuWa08}, which arises in the weak-coupling limit.
By contrast, we investigate a strong-coupling regime in which the
inhibitory pulses can prevent individual neurons from firing.

As a main result we find that different $\gamma$-rhythms can become
synchronized by noise even if that noise consists of independent Poisson
spike trains and is therefore completely uncorrelated between different
neurons and networks. This is in contrast to the well-studied case
of noise-induced synchronization for which noise correlations are
essential \cite{Pi84,TeTa04}. By reducing the coherent dynamics of
the two networks to a minimal iterated map we show that the noise-induced
phase heterogeneity allows the faster network to suppress the spiking
of a fraction of the neurons in the slower network. This increases
the frequency of the slower network and allows it to synchronize with
the faster network. The synchronization leads to a more consistent
phase relationship in the output of the two networks. We illustrate
that this can increase the learning speed of downstream neurons that
read the network output via synapses exhibiting spike-timing-dependent
plasticity.

We consider two coupled networks of $N/2$ integrate-fire neurons
each that receive $C_{1}=\epsilon_{1}N$ random inhibitory connections
from their own network and $C_{2}=\epsilon_{2}N$ random inhibitory
connections from the other network. Thus, all neurons have the same
in-degree. In addition, each neuron receives noisy external excitatory
inputs $I_{i}^{(ext)}(t)$ . The depolarization $V_{i}(t)$ of neuron
$i$, $i=1,\dots,N$, is given by 
\begin{equation}
\tau\dot{V}_{i}=V_{rest}-V_{i}+RI_{i}^{(syn)}(t)+RI_{i}^{(ext)}(t),\label{eq:kirchhoff}
\end{equation}
where $I_{i}^{(syn)}(t)$ denotes the total synaptic current from
within the networks, $\tau$ the membrane time constant, and $R$
the membrane resistance. When $V_{i}(t)$ reaches the firing threshold
$V_{\theta}$, a spike is triggered and the voltage is reset to the
reset voltage $V_{r}$.

The synaptic currents are modeled as the difference of two exponentials,
triggered by spikes of the presynaptic neuron $j$ at times $t_{j}^{(k)}$,
\[
I_{i}^{(syn)}=\frac{g_{syn}}{R}\left(A_{i}^{(2)}-A_{i}^{(1)}\right)\left(V^{(rev)}-V\right),
\]
with
\begin{equation}
\dot{A}_{i}^{(1,2)}=-\frac{A_{i}^{(1,2)}}{\tau_{1,2}}+\sum_{j=1}^{N}\sum_{k}W_{ij}\,\delta\left(t-t_{j}^{(k)}-\tau_{d}\right).
\end{equation}
Here $V_{i}^{(rev)}$ denotes the reversal potential, $g_{syn}$ the
dimensionless synaptic strength, and $\tau_{d}$ the synaptic delay.
The connectivity matrix is denoted by $\mathbf{W}$ with its non-zero
elements given by $W_{ij}=1$ if neuron$i$ and $j$ belong to the
same network while $W_{ij}=\gamma_{0}$ if they belong to different
networks. 

The external input of each neuron $i$ is modeled as an independent
Poisson-train of $\delta$-spikes at times $t_{ik}^{(ext)}$, 
\begin{equation}
I_{i}^{(ext)}(t)=\frac{1}{R}\Delta V_{i}\sum_{k}\delta(t-t_{ik}^{(ext)}).
\end{equation}
Thus, the noisy external inputs to different neurons are uncorrelated.
The dimensionless input strengths $\Delta v_{i}\equiv\Delta V_{i}/(V_{\theta}-V_{r})$
are equal for all neurons within a network, but may differ between
the two networks: $\Delta v_{i}=\Delta v^{(l)}$ for neurons in network
$l$ with $\Delta v^{(2)}=\rho\Delta v^{(1)}$. Instead of the spike
rates $\lambda$ and the strengths $\Delta v^{(1)}$ we use the mean
input $\mu=\lambda\Delta v^{(1)}$and its variance $\sigma^{2}=\lambda(\Delta v^{(1)})^{2}$
as independent parameters.

\begin{figure}
\centering{} \includegraphics[width=1.05\linewidth]{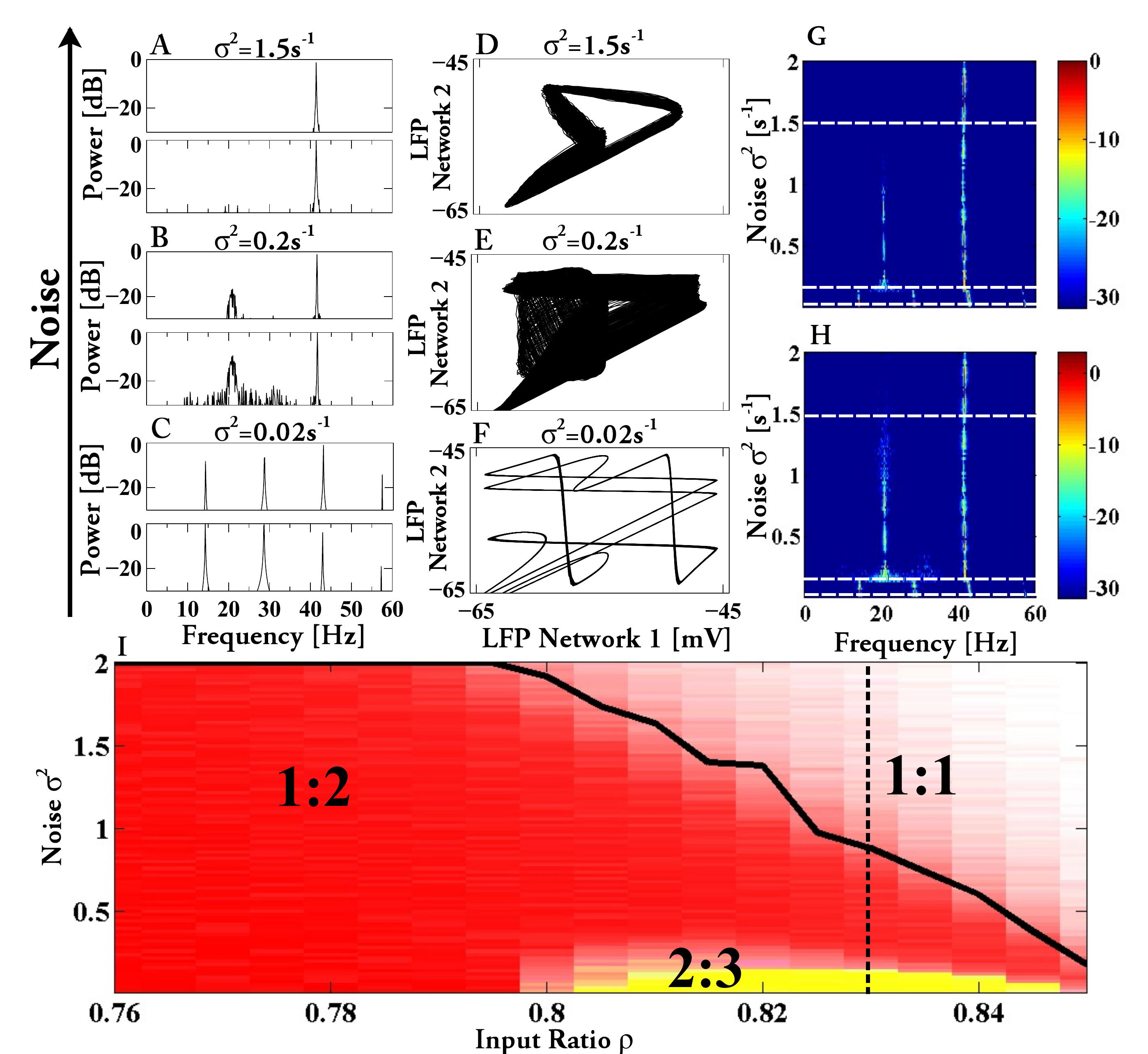}
\caption{Independent noise synchronizes population activity of two coupled
networks (cf. Fig.\ref{fig:figure1}A). Fourier spectra of the two
networks (A,C,E correspond to the dashed lines in G,H) and the corresponding
attractors (B,D,F) for $\rho=0.83$. Phase diagram showing transitions
between different phase-locked and synchronized states as a function
of noise. Color hue indicates frequency ratio, intensity the logarithmic
power of the dominant Fourier mode (I). Parameters: $N=1,000$, $\epsilon_{1}=0.28$,
$\epsilon_{2}=0.12$, $\tau=20\mbox{ms}$, $\tau_{1}=4.82\mbox{ms}$,
$\tau_{2}=5.37\mbox{ms}$, $\tau_{d}=2\mbox{ms}$, $V_{rest}=-55\mbox{mV}$,
$V_{\theta}=-45\mbox{mV}$, $V_{r}=-65\mbox{mV}$, $V^{(rev)}=-85\mbox{mV}$,
$g_{syn}=0.015$, $\gamma_{0}=1.5$, $\mu=200s^{-1}$.\textbf{ }}
\label{fig:figure2} 
\end{figure}
Due to their inhibitory coupling the neurons within each network synchronize,
resulting in a population rhythm, which corresponds to an interneuronal
network $\gamma$-rhythm (ING) \cite{TiSe09}. We characterize it
here via the population mean $V_{mean}^{(l)}$ of the voltage as a
proxy for the local field potential (LFP).

Synchronization is usually achieved by increasing the coupling strength,
while noise tends to decrease the degree of synchronization. However,
for the two coupled ING-rhythms our numerical simulations show that
increasing the strength $\sigma^{2}$ of the independent noise - at
fixed coupling strength - can enhance the synchrony of the networks
(Fig.\ref{fig:figure2}). While adding small amounts of noise smears
out the attractor, here represented in terms of the LFPs of the two
networks (Fig.\ref{fig:figure2}D,F), stronger noise `cleans up' the
attractor (Fig.\ref{fig:figure2}B), which is reflected in a reduction
of the low-frequency components of the Fourier spectra (Fig.\ref{fig:figure2}A,C,E).
Figures \ref{fig:figure2}G,H show the spectral power of the two networks
as a function of noise in terms of a logarithmic colorscale. The dashed
lines indicate the values of the noise used in Fig.\ref{fig:figure2}A-F.

As the ratio $\rho$ of the mean inputs is changed, the frequency
ratio of the rhythms changes. This leads to domains akin to Arnold
tongues. For intermediate values of $\rho$ and low noise the two
LFPs show phase-locked behavior with a frequency ratio of $2:3$ (Fig.\ref{fig:figure2}I).
As the noise is increased a transition to a ratio of $1:2$ is found,
with the subharmonic response fading away in an inverse period-doubling
bifurcation as the noise is increased further. For lower $\rho$ the
$1:2$-tongue arises without noise. It also undergoes an inverse period-doubling
bifurcation with increasing noise.

\begin{figure}
\centering{}\includegraphics[width=1.05\linewidth]{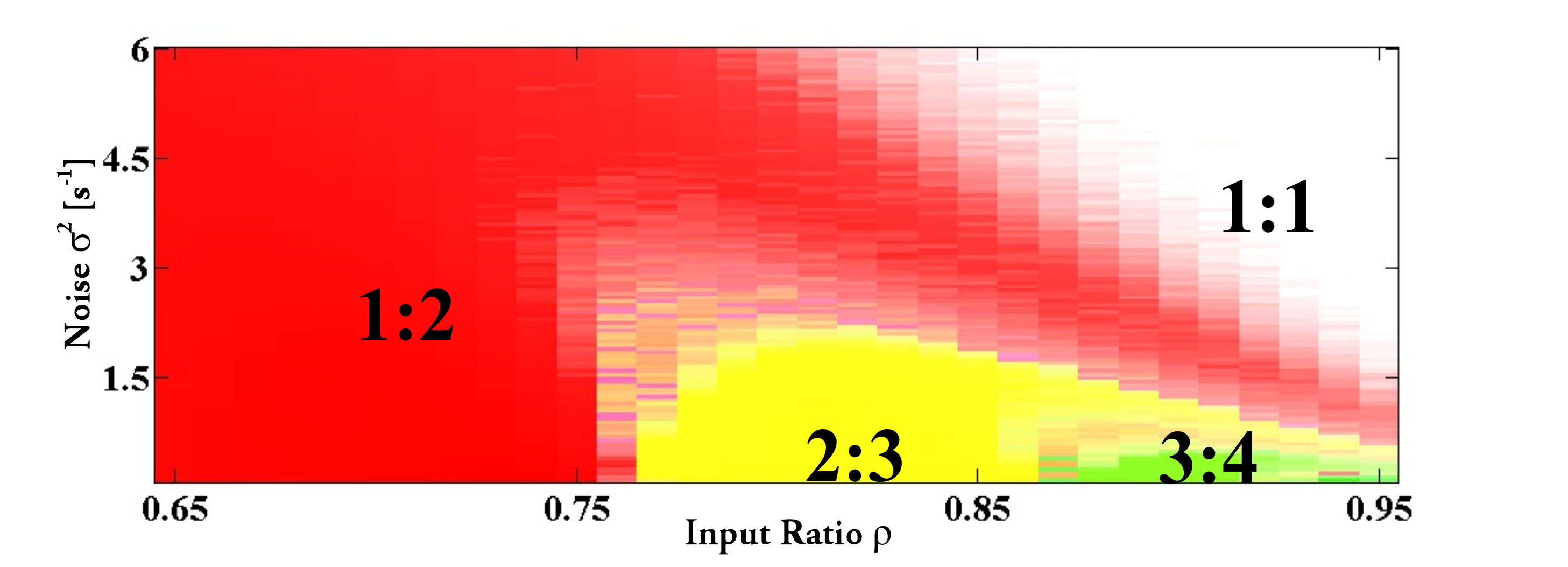}
\caption{Phase diagram for a single network with periodic inhibition. Each
neuron in the network receives independent noise. Parameters and colors
as in Fig.\ref{fig:figure2}I. }
\label{fig:figure3} 
\end{figure}

To further our understanding we consider the impact of strictly periodic
inhibition on the slower network 2 (Fig.\ref{fig:figure1}B). We generate
that inhibition using network 1, omitting its inhibition by network
2 and using noiseless input, which is 20\% reduced to compensate for
the reduced inhibition. The behavior of this simplified system is
qualitatively very similar to that of the bidirectionally coupled
networks (Fig.\ref{fig:figure3}). Again, as the noise is increased
the system undergoes transitions between different phase-locked states
and eventually reaches the synchronized $1:1$-state via an inverse
period-doubling bifurcation.

\begin{figure}
\centering{}\includegraphics[width=1\linewidth]{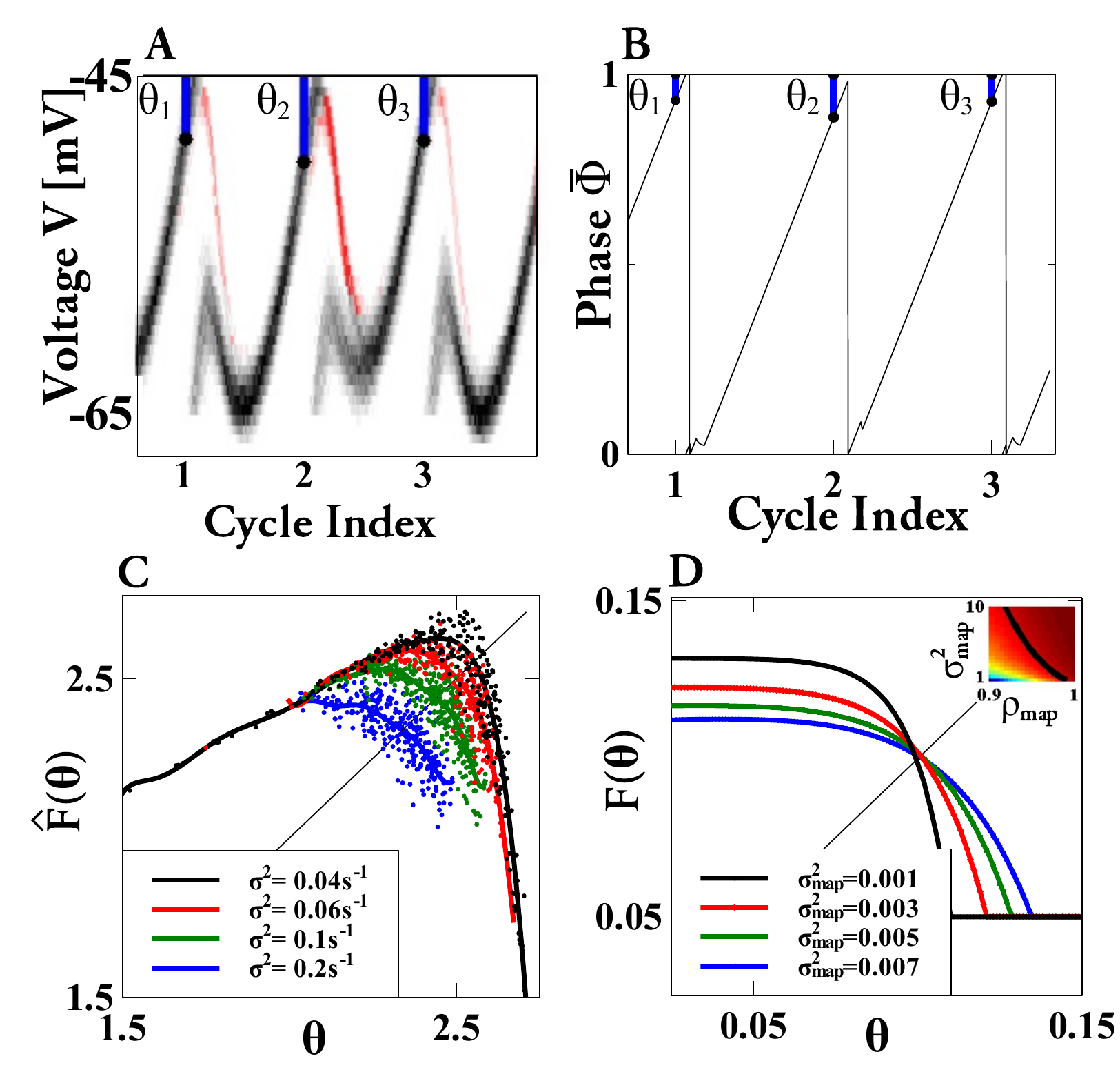}
\caption{Network simulations (A,C) compared with results of the minimal map
(B,D). Time-dependence of the voltage distribution (A) and of the
mean phase $\bar{\phi}$ (B) and the lag $\theta$ of the activity
of network 2. C) Iterated map extracted from direct simulations..
D) Iterated map from (\ref{eq:phase change 2}) and its slope at the
fixed point (inset)\textbf{. }Period-doubling bifurcation marked by
line in inset. Parameters: (A,C) as in Fig.\ref{fig:figure3} except
$\rho=0.92$, $\sigma^{2}=1.8$ in (A) and $\rho=1.05$,$\gamma_{0}=1.9$
in (C); (B,D) \textbf{$e^{-g_{1}}=0,$$e^{-g_{2}}=0.1,$$\tau_{d}=0.1$,
$\rho_{map}=0.95$. }}
\label{fig:figure4} 
\end{figure}

The temporal evolution of the voltage distribution of the neurons
in network 2 provides insight into the synchronization mechanism (Figure
\ref{fig:figure4}A, darker color denotes larger number of neurons
with that voltage). Not all neurons spike in each cycle: some do not
reach the threshold before the periodic external inhibition sets in
and keeps them from spiking (marked red in Fig.\ref{fig:figure4}A).
Their voltage decreases smoothly instead of the instantaneous reset
to $V_{r}$ after a spike. The fraction of neurons that spike varies
from cycle to cycle. In the regime of interest the two peaks of the
voltage distribution that correspond to spiking and non-spiking neurons
are pushed together by the strong inhibition (Figure \ref{fig:figure4}A)
and before the neurons in network 2 spike again and before the next
cycle of the periodic inhibition sets in the voltage distribution
becomes unimodal. This allows us to develop a phenomenological Poincaré
map for the mean phase $\bar{\phi}$ of the neurons. Assuming that
the inhibition resets the phase of an oscillator by an amount proportional
to the phase, we write the evolution of the mean phase as 
\begin{equation}
\dot{\bar{\phi}}=-g_{1}P_{1}(t)\,\bar{\phi}(t)-g_{2}P_{2}(\bar{\phi}(t-\tau_{d}))\bar{\phi}(t)+\rho_{map},\label{eq:mean phase evolution}
\end{equation}
with $\bar{\phi}$ being reset to $\bar{\phi}=0$ instantaneously
when it reaches $\bar{\phi}=1$. The first term in (\ref{eq:mean phase evolution})
represents the periodic external forcing with strength $g_{1}$ and
frequency 1. The second term models the self-inhibition of the network,
which arises from those oscillators that are at the spike threshold
when the average phase has the value $\bar{\phi}$. Their number is
denoted by $P_{2}(\bar{\phi}(t))$. It reflects the phase distribution
of the oscillators and the resulting heterogeneity in the spiking
times. The simulations indicate that this heterogeneity plays a central
role (Fig.\ref{fig:figure4}A). Instead of considering an evolution
equation for this distribution, for our minimal model we consider
it time-independent and of the form
\begin{equation}
P_{2}(\bar{\phi})=\left\{ \begin{array}{cc}
\frac{1}{\sigma_{map}} & \qquad\bar{\phi}\in[0,\frac{1}{2}\sigma_{map}]\cup[1-\frac{1}{2}\sigma_{map},1)\\
0 & \mbox{otherwise.}
\end{array}\right.\label{eq:phase distribution}
\end{equation}
Thus, for $\bar{\phi}\in[0,\sigma_{map}/2]$ neurons in the trailing
half of the distribution are firing, while for $\bar{\phi}\in[1-\sigma_{map}/2,1)$
neurons in the leading half are firing.

Letting $t_{k}^{(2)}$ be the time at which the mean phase $\bar{\phi}$
reaches threshold, we focus on the situation in which the external
inhibition arrives before any of the self-inhibition triggered by
the oscillators in network 2 sets in, $t_{k}^{(1)}+\tau_{d}<t_{k}^{(2)}-\sigma_{map}/\left(2\rho_{map}\right)+\tau_{d}$.
The external inhibition induces a phase reset $\bar{\phi}(t_{k}^{(1)}+\tau_{d})\rightarrow e^{-g_{1}}\bar{\phi}(t_{k}^{(1)}+\tau_{d})$.
For sufficiently strong coupling $g_{1}$ it keeps the trailing oscillators
from spiking and from contributing to the self-inhibition. Thus, self-inhibition
lasts from $t_{k}^{<}=t_{k}^{(2)}-\sigma_{map}/2$ to $t_{k}^{>}=\min(t_{k}^{(1)}+\tau_{d},t_{k}^{(2)}+\sigma_{map}/\left(2\rho_{map}\right)+\tau_{d})$.
During that time $\Delta t$ it induces a phase change that leads
to
\begin{equation}
\bar{\phi}(t_{k}^{>})=e^{-g_{2}\frac{\rho_{map}\,\Delta t}{\sigma_{map}}}\phi(t_{k}^{<})+\frac{\sigma_{map}}{g_{2}}(1-e{}^{-g_{2}\frac{\rho_{map}\,\Delta t}{\sigma_{map}}}).\label{eq:phase change 2}
\end{equation}

Combining (\ref{eq:phase change 2}) with the phase evolution during
the remaining time yields a Poincaré map for the phase lag $\theta_{k}\equiv1-\bar{\phi}\left(t_{k}^{(1)}\right)$
of network 2 relative to the periodic inhibitory input, $\theta_{k+1}=F(\theta_{k})$
(Fig.\ref{fig:figure4}B,D). The fixed point $\theta_{FP}=F(\theta_{FP})$
corresponds to a 1:1 synchronized state. It is only stable for large
widths $\sigma_{map}$ of the distribution $P_{2}$, i.e. for sufficiently
strong noise, and becomes unstable via a period-doubling bifurcation
as the noise is reduced, capturing a striking, common feature of the
full network simulations (Figs. \ref{fig:figure2}D,\ref{fig:figure3}D). 

The phase lag of network 2 relative to the periodic forcing extracted
from the full simulations yields a noisy map that also undergoes a
period-doubling bifurcation as the noise is increased (Fig.\ref{fig:figure4}C).
Also captured by the phenomenological map is the simulation result
that the noise level needed to stabilize the synchronized state increases
with decreasing $\rho$, i.e. with increasing difference in the frequencies
of the uncoupled networks (inset Fig.\ref{fig:figure4}D).

The minimal model (\ref{eq:mean phase evolution},\ref{eq:phase distribution})
identifies as a key element of the synchronization the phase-dependent
spiking fraction of network 2. Even though all neurons in this network
receive the same mean external input, the noise in that input induces
a spread in the phase. This allows the strong inhibition from network
1 or from the periodic forcing to suppress the spiking of the neurons
that happen to be in the tail of the phase distribution, while the
neurons in the front escape that inhibition. With fewer neurons spiking,
the total self-inhibition within network 2 is reduced, speeding up
the rhythm of network 2 in the following cycle. If the rhythm of network
2 is faster, more of its neurons spike, increasing self-inhibition
and slowing down the rhythm. If this stabilizing feedback is removed
by adjusting in each cycle the strength of the self-inhibition to
compensate for the variable fraction $f_{s}$ of spiking neurons,
$G_{syn}\rightarrow G_{syn}/f_{s}$, synchronization is lost (data
not shown).

What functional consequences may the synchronization of rhythms by
independent noise entail? Since it renders the spike timing of the
neurons in network 2 more consistent relative to the spikes in network
1, it may, for instance, impact learning processes that are based
on spike-timing dependent plasticity (STDP), an essential component
of learning in the nervous system. We therefore consider a read-out
neuron that is driven by both networks via synapses exhibiting STDP.
To obtain the required excitatory outputs in a biologically convincing
fashion our minimal inhibitory network would have to be extended to
include also excitatory neurons \cite{BoKo03}. For simplicity we
assume here that the spiking of the excitatory neurons in such an
E-I network is sufficiently tightly correlated with the spiking of
the inhibitory neurons that we can take the output of the inhibitory
neurons as a proxy for the excitatory output.

\begin{figure}
\centering{}\includegraphics[width=1\linewidth]{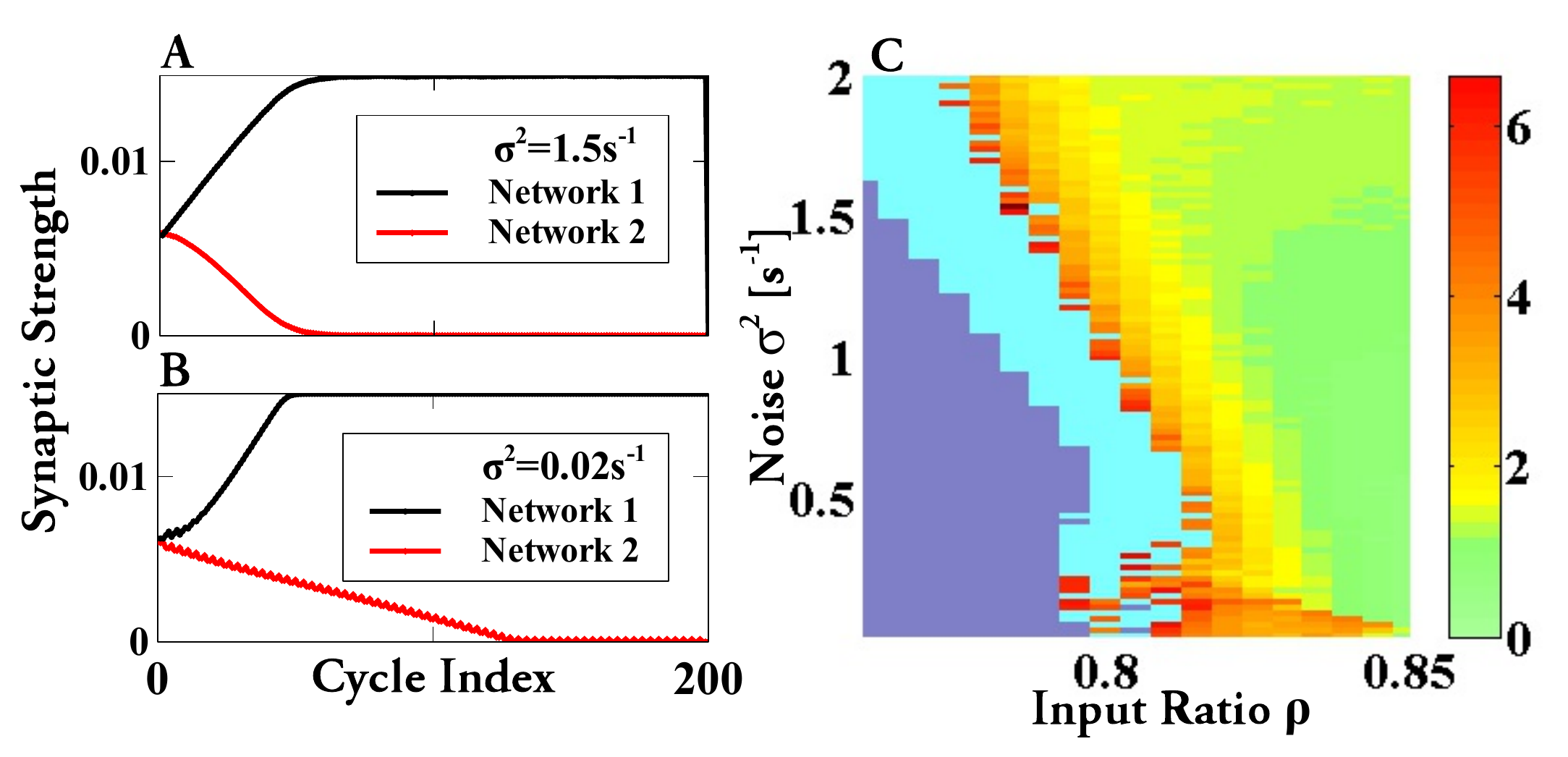}
\caption{Learning is accelerated by independent noise. (A,B) Temporal evolution
of the mean synaptic weight of the output synapses of network 1 and
2 with and without noise. (C) Learning duration $T_{l}$ as a function
of noise and input ratio $\rho$. Light Blue: $T_{l}>6.7s$. Purple\textbf{
}: synaptic weights increase rather than decrease. Parameters: $A_{p}=A_{m}=3\cdot10^{-5}$,
$\tau_{p}=\tau_{m}=5\mbox{ms}$ and as in Fig.\ref{fig:figure2}. }
\label{fig:figure6} 
\end{figure}

For the plastic synapses we take the STDP rule \cite{BiPo98}
\begin{equation}
\Delta G=\left\{ \begin{aligned}A_{p}e^{-(t_{post}-t_{pre})/\tau_{p}} & \mbox{ for } & t_{post}-t_{pre}>0\\
-A_{m}e^{(t_{post}-t_{pre})/\tau_{m}} & \mbox{ for } & t_{post}-t_{pre}<0
\end{aligned}
\right.
\end{equation}
where $\Delta G$ is the change in the synaptic weight $G_{syn}$
when the read-out neuron spikes at $t_{post}$. The spike time of
the respective pre-synaptic neuron is denoted by $t_{pre}$. For each
presynaptic neuron each postsynaptic spike is paired with only adjacent
presynaptic spikes. The weights $G_{syn}$ are restricted to $0\le G_{syn}\le0.015$. 

When two networks with slightly different intrinsic frequencies are
coupled, the spikes of the faster network consistently precede the
spikes of the read-out neuron, independent of the degree of synchronization
of the two rhythms. This induces a monotonic increase in the weights
of the output synapses of the faster network (Fig.\ref{fig:figure6}).
This is, however, not the case for the slower network. Without synchronization
and after a transient some of its spikes arrive after and some before
the spikes of the read-out neuron. The resulting non-monotonic change
in the synaptic weights slows down their overall decay (Fig.\ref{fig:figure6}A).
However, when the rhythms of the two networks are synchronized by
independent noise the neurons in the slower network always spike after
the read-out neuron, resulting in a much faster decay of their synaptic
weights. Thus, the synchronization by independent noise can enhance
the speed with which read-out neurons select between different input
networks.

The learning duration, defined as the time it takes for the mean of
the output weights of the slower network to decay to $3\cdot10^{-4}$,
reflects in detail the tongue structure of the phase diagram (compare
Fig.\ref{fig:figure6}B with Fig.\ref{fig:figure2}I). A more detailed
inspection shows that near the period-doubling bifurcation from the
$1:1$-state to the $1:2$-state the learning duration exhibits a
minimum (line in Fig.\ref{fig:figure2}I): while in the $1:1$-state
a decrease in noise reduces the variability in the spike timing, a
similar decrease in noise in the $1:2$-state increases the variability
because it predominantly increases the period-doubling amplitude.

In conclusion, we have shown that independent noise can synchronize
population rhythms of coupled inhibitory networks of spiking neurons.
As the underlying mechanism we have identified the phase heterogeneity
of the neurons that results from the noise, which allows the faster
network to suppress the spiking of a fraction of the neurons in the
slower network. Since this mechanism requires variability in the spiking
fraction it can only operate in networks; individual neurons are not
synchronized by the uncorrelated noise. In fact, when the network
size is reduced below $\mathcal{O}(100)$ neurons, the synchronization
deteriorates significantly, because silencing an individual neuron
impacts the self-inhibition too much. 

The nature of the synchronization mechanism suggests that heterogeneity
of neuronal properties instead of noisy inputs should similarly enhance
the synchronizability of population rhythms in coupled networks. Preliminary
simulations indicate this to be the case.

We gratefully acknowledge support by grant NSF-CMMI1435358. This research
was supported in part through the computational resources and staff
contributions provided for the Quest high performance computing facility
at Northwestern University.

\bibliographystyle{unsrt}
\bibliography{../../Databases/journal}

\end{document}